# Broken particle-hole symmetry at atomically flat *a*-axis YBa$_2$Cu$_3$O$_{7-\delta}$ interfaces

Bruce A. Davidson[1, 2, a], Revaz Ramazashvili[1, 3], Simon Kos[1, 4,b], and James N. Eckstein[1]

[1]*Physics Department, University of Illinois at Urbana-Champaign, Urbana, IL 61801, USA.*

[2]*INFM-TASC National Laboratory, Area Science Park, 34012 Basovizza (TS), Italy.*

[3]*Materials Science Division, Argonne National Laboratory, Argonne, IL 60439, USA.*

[4]*Center for Nonlinear Studies, Los Alamos National Laboratory, Los Alamos, NM 87545, USA.*

We have studied quasiparticle tunneling into atomically flat *a*-axis films of YBa$_2$Cu$_3$O$_{7-\delta}$ and DyBa$_2$Cu$_3$O$_{7-\delta}$ through epitaxial CaTiO$_3$ barriers. The junction heterostructures were grown by oxide molecular beam epitaxy and were carefully optimized using *in-situ* monitoring techniques, resulting in an unprecedented crystalline perfection of the superconductor/insulator interface. Below $T_c$, the tunneling conductance shows the evolution of a large unexpected asymmetrical feature near zero bias. This is evidence that superconducting YBCO crystals, atomically truncated along the lobe direction with a titanate layer, have intrinsically broken particle-hole symmetry over macroscopically large areas.

---

[a] Corresponding author's email: davidson@tasc.infm.it

[b] Current address: Dept. of Physics, Univ. of Cambridge, Cambridge CB3 0HE, UK.

Tunneling into superconductors probes the electronic structure of quasiparticle (qp) excitations. According to the Bogoliubov description in conventional BCS theory, at the Fermi energy qps are made up of electrons and holes with equal weight. Tunneling electrons into or out of the superconductor reveals the hole and particle content of the excitations, and in the absence of magnetism produces particle/hole (p/h) symmetric conductance spectra. For this reason, locally broken p-h symmetry in the qp density of states (DoS) seen near isolated impurities [1,2] and vortices [3,4] in cuprate superconductors has provided some of the most detailed experimental data on order parameter symmetry and intrinsic magnetism in these materials [5,6]. Experimentally, these impurity-induced DoS perturbations behave counter-intuitively; e. g., studies of the local DoS near nominally nonmagnetic Zn in $Bi_2Sr_2CaCu_2O_{8+\delta}$ [1] have been explained by invoking the Kondo effect [7,8] while nominally magnetic Ni causes less local order parameter disruption than Zn [2]. Similarly, chain oxygen vacancies in $YBa_2Cu_3O_{7-\delta}$ (YBCO) cause strong resonances in the local DoS that are distributed in energy throughout the gap [9] and indicate the magnetic character of the vacancy-Cu complex. These experiments play a crucial role in theoretical discussions of cuprates [10,11] and point to the utility of studying the electronic behavior of new defect structures. They have been possible only on the *c*-axis surfaces that result from cleaving single crystals (necessarily in UHV and below 20 K). Investigation of the DoS of other surfaces (such as an abruptly truncated (100) crystal) and other defect structures (such as a linear impurity array) requires either *in-situ* vacuum tunneling into the as-grown surface or epitaxial stabilization of the defect structure (e.g., in the form of an interface) by a suitable tunnel barrier.

In this paper, we report p-h asymmetric conductance spectra from tunneling into an atomically flat (100) YBCO surface through a nonmagnetic insulator, and we propose this asymmetry is caused by the underlying magnetism of the cuprate states. The tunnel barrier is a thin heteroepitaxial $CaTiO_3$ film, and Fig. 1 shows the arrangement of Cu-O planes and chains at the interface in this geometry; it is the precise truncation of the lattice at this plane that leads to our new results. We find a large particle-hole asymmetry in the subgap DoS in the superconducting state. We attribute these modifications of the DoS to qp scattering at the interface which behaves as a linear array of impurities. We have measured both the temperature and doping dependence of the asymmetry, and find it is strongest at optimal doping and weakens with underdoping, disappearing as the critical temperature $T_c \rightarrow 50K$.

The tunnel junctions were fabricated from NIS trilayer films grown by ozone assisted molecular beam epitaxy, which has demonstrated excellent control of atomic layering of transition metal perovskite materials [12]. We used *in-situ* RHEED to characterize the surface crystallography and morphology during the film and heteroepitaxial barrier growth. Pure (100) orientation of 200 nm thick YBCO and DyBCO films is accomplished by the self-template technique [13] on carefully prepared (100) $SrTiO_3$ substrates [14]. After YBCO deposition [Fig. 2(a)], the RHEED specular spot (SS) intensity was equal to that of the starting substrate, and the $1/3^{rd}$- and $1^{st}$-order spots were comparable in intensity, with no reconstruction or other spots visible. These characteristics indicate that the surface was very close to an ideally truncated (100) crystal. *Ex-situ* atomic force microscopy of YBCO films grown with identical RHEED characteristics showed RMS roughness $< 4 Å$ over many $mm^2$, and asymmetric x-ray diffraction reciprocal space maps showed pure *a*-axis orientation and pseudomorphic growth up to 200 nm [15].

Directly following the cuprate growth, an insulating barrier of precisely 4, 5 or 6 unit cells of CaTiO$_3$ (CTO) was deposited. The CTO grew two-dimensionally and pseudomorphically, showing one SS oscillation/monolayer. The 1/3$^{rd}$-order spots disappeared within 2 CTO layers [Fig. 2(b,c)], and the SS was completely recovered by 5 layers [Fig. 2(d)]. Taken together, these features indicate uniform barrier coverage. Doping was controlled via a predetermined cooling procedure [16], followed by *in-situ* Au counterelectrode deposition.

The NIS trilayer was processed into vertical mesa junctions with cross-sections $50-1000$ μm$^2$. For this study, we fabricated junctions ranging from near optimal ($T_c = 83$ K) to moderately underdoped ($T_c = 55$ K). We note that surface $T_c$, as determined by the initial appearance of the gap structure described below, and bulk $T_c$, determined by transport measurements on the film directly underneath, agree to within a few degrees. This is strong supporting evidence that the (100) surface sampled in the tunneling experiments is not marred by extrinsic effects (interface disorder or oxygen loss), and that the drastically altered superconductivity seen in qp DoS is entirely due to the abrupt (100) termination.

Differential conductance $G(V)$ measurements identify tunneling as the dominant transport mechanism. $G(0)$ has a weakly insulating $T$ dependence from 300 to 100 K (decreasing less than a factor of 2), and below $T_c$, $G(100$ mV$)$ is nearly $T$-independent. Moreover, $G(0)$ scales linearly with junction area and exponentially with barrier thickness (a factor of 12 for each titanate unit cell, at fixed doping). In addition, well into the normal state, $G(V)$ has the general form of a shifted parabola with minimum at $\approx +20$ mV for all junctions [Fig. 3]. This is consistent with elastic tunneling between metals of different work functions, resulting in a trapezoidal barrier profile [17]. Estimates of barrier thickness and heights from fitting $G(V)$ at 100 K as in Ref. [17] are reasonable compared to other perovskite junctions with titanate barriers we have grown [18]. Fit barrier thicknesses are within 25% of the nominal, and barrier heights for electron tunneling in the ~40 junctions included in this study are $800 \pm 250$ meV (YBCO) and $205 \pm 120$ meV (Au).

The most important new result reported here is the unusual low bias ($|V| \leq 50$ mV) behavior of $G(V)$ below $T_c$. At negative bias, corresponding to tunneling of qps from cuprate states with mostly particle-like character ("occupied" states) into empty Au states, $G(V)$ acquires a gap-like feature that reflects the onset of superconductivity. At $T \ll T_c$, near optimal doping the occupied state DoS [Figs. 3(a) and 4, normalized] show a coherence peak near $-25$ meV whose energy is independent of $T$, and a "peak-dip-hump" feature. This behavior is in good quantitative agreement with angle-resolved photoemission data on untwinned, optimally doped $c$-axis YBCO at the $(\pi, 0)$ point in the Brillouin zone [19]. STM/S on optimally doped $c$-axis YBCO has also shown a $T$–independent coherence peak energy [20]. Integrating the normalized $\int_{-60mV}^{0} G(T,V) dV$, DoS is conserved as expected (to normalization error, $\leq 2\%$).

In stark contrast, $G(V)$ for positive bias shows an unexpected and dramatic step-like increase just above $V = 0$ where the "gap" is expected by symmetry. This bias corresponds to electrons tunneling from the gold into states with mostly hole-like character in the cuprate. For $V > 0$, the integrated "empty" state DoS is not conserved; the data near optimal doping [Fig. 4] shows 12% more states at low $T$ as compared to the normal state. This clearly demonstrates the breaking of symmetry between particle-like and hole-like DoS at this interface in the superconducting state.

The extra spectral weight has a roughly $T$-linear dependence (not shown), vanishing at $T_c$. For samples with lower $T_c$ the DoS asymmetry is less pronounced [Fig. 3(b, c)]. The extra spectral weight can be used to quantify the strength of symmetry breaking for different dopings, and is summarized for all junctions [Fig. 4, inset]. We emphasize that *every* epitaxial junction with pure *a*-axis orientation we have measured shows extra states only above the chemical potential and inside the gap, with occupied state DoS conserved. The $G(V)$ curves are also relatively insensitive to magnetic fields up to 7 Tesla (parallel or perpendicular to the surface) at any voltage, including low bias. This rules out an explanation involving collectively generated interfacial currents, such as when the splitting of a zero bias conductance peak increases in field [21].

Thus, we conclude that the step-like asymmetry at $V = 0$ is a robust feature of this (100) interface, and the strength of the asymmetry is correlated with doping. This implies that YBCO crystals atomically truncated along the lobe direction by an epitaxial CTO layer have intrinsic p-h asymmetry in the superconducting state. It is in marked contrast to all tunneling data on *c*-axis, (110) or rough (100) surfaces, with or without impurities or defects, in that here the p-h asymmetry is a *global* (averaged) property of the atomically flat surface. We contrast our results with symmetric DoS seen in the $G(V)$ of (110) and rough (100) YBCO surfaces [22] that show a pronounced zero-bias conductance peak (ZBCP) attributed to Andreev bound states (ABS) [23]. For flat lobe-facing surfaces such as (100), the ABS model predicts a p-h symmetric DoS with no ZBCP. Flat (100) YBCO surfaces with amorphous barriers have shown a ZBCP [24] attributed to the nonspecular interface. We also contrast our results with small asymmetric sloping backgrounds in STM spectra of *c*-axis Bi-2212 outside the gap [25] in which the coherence peaks and subgap $G$ are nearly symmetric about $V = 0$.

There are two aspects of this data that deserve particular attention: (i) the low bias asymmetry appearing only in the superconducting DoS, with its unusual doping dependence, and (ii) the large residual subgap conductance evident in $G(V)$: even at $T \ll T_c$, the subgap conductance never vanishes, but instead has a minimum of ~80% of $G(0)|_{T=T_c}$. Concerning (ii), we note that all planar cuprate NIS junctions in the literature (*a-b* plane and *c*-axis) show a similar or larger subgap $G$ [26], and its origin is not understood. It is usually attributed to weakened superconductivity at the surface, either intrinsically or due to disorder at the interface. In our case it persists in spite of the fact that the interface epitaxy has been carefully optimized and that the tunneling (with a narrow forward cone estimated $\leq 25°$) probes qp states near antinodes where the energy gap is fully formed and $G(0)|_{T=0}$ should approach zero.

Regarding point (i), two qualitative scenarios based on the way in which plane and chain DoS are perturbed by local defects may be relevant to our data. The low bias asymmetry of our data may be compared to asymmetric resonances observed in the local DoS near isolated Ni impurities [2] or individual YBCO chains [9], both of which may show substantial p-h asymmetry, but yield nearly symmetric DoS when averaged over large ($>100\text{x}100$ Å$^2$) areas [27]. That is, extra spectral weight in empty-state DoS at one site is compensated for by extra weight in occupied-state DoS at other sites. In our geometry, the heteroepitaxial interface consists of nearly ideal one-dimensional $CuO_2$ plane edges, as well as interleaved $CuO_{1-\delta}$ chains [Fig. 1]. The truncation of the $CuO_2$ plane with a row of $TiO_3$ octahedra may perturb the qp DoS at the Cu site along the plane edge in a manner analogous to that seen experimentally [2] at atomic sites around single Ni defects and predicted theoretically [28,29]. It is precisely into these sites that tunneling occurs and, because of the translational symmetry of the interface (all sites are equivalent), the asymmetry in the tunneling DoS would no longer disappear upon averaging over the interface. This interpretation, while offering no explanation of the large residual subgap $G$, suggests that p-h asymmetry is a generic feature of flat, epitaxial (100) cuprate-titanate interfaces, and could be corroborated by STM/S studies of individual Ti impurities substituted on Cu plane sites (which to our knowledge have not yet been done).

Similarly, a perturbation of the chain-derived DoS at the interface may arise from a different oxygen coordination of chain copper atoms next to the CTO layer. In unperturbed chains, $s$–wave pairing is induced by proximity to superconducting planes, and p-h asymmetric resonances localized to the chains in YBCO$_{6.97}$ as seen by STM/S [9] imply a magnetic character to the scattering defects. A theory of these resonances [11] may be relevant here, assuming that perturbed (surface) chains remain superconducting by proximity in the $a$-axis geometry, and that they represent a tunneling channel comparable in strength to plane edges. Scattering resonances in this model may contribute to a large residual subgap $G$, or to a p-h asymmetric DoS, or both, depending on the sign and distribution of the scalar component $U_0$ of the impurity potential (notation of Ref. [11]) that characterizes the defects. A distribution involving large $U_0$ would lead to nonzero residual $G$. If all defects are identical (e.g., chain O vacancies), they are likely to produce $U_0$ of a definite sign, and thus lead to p-h asymmetry on top of residual $G$. This could explain the doping dependence of both the residual subgap $G$ and p-h asymmetry seen here: as the defect density increases, bound states begin to interfere, gradually washing out the asymmetry and further filling in the gap [30]. Since our $G(V)$ curves are an average over the (macroscopic) interface area, we cannot determine whether one or both of these DoS perturbation mechanisms (plane edges or chains) are relevant. A microscopic tunneling study of this interface configuration, or analogous studies of qp tunneling in epitaxial junctions using a chainless cuprate, may resolve this issue.

To conclude, we have carried out the first systematic study of *a*-axis tunneling into atomically flat YBCO with an epitaxial titanate barrier. We observe a remarkable subgap asymmetry of qp DoS in the superconducting state, which we interpret as a consequence of the termination of the cuprate crystal at the boundary. Since the interface is regular and homogeneous, this result provides a clear test for theories of high Tc superconductivity.

We thank K. E. Gray, D. K. Morr, M. R. Norman, M. V. Klein, P. J. Hirschfeld and I. Vekhter for discussions, and E. Colla for technical assistance. B. A. D. and R. R. thank the Condensed Matter Theory group at the Abdus Salam International Center for Theoretical Physics (ICTP, Trieste) for kind hospitality. B. A. D. and J. N. E. were supported by US ONR grant #N00014-00-1-0840; R. R. by U.S. DOE, Basic Energy Sciences–Materials Science, contract #W-31-109-ENG-38 and, in part, the MacArthur Foundation at UIUC;
S.K. by contract #NSF-DMR-98-1794 while at UIUC and LANL DR Project

200153 and U.S. DOE contract #W-7405-ENG-36 while at LANL. We acknowledge use of the F. Seitz MRL-CMM and MRL-Microfabrication Lab (UIUC) through U.S. DOE, Div. of Materials Science award #DEFG02-91ER45439.

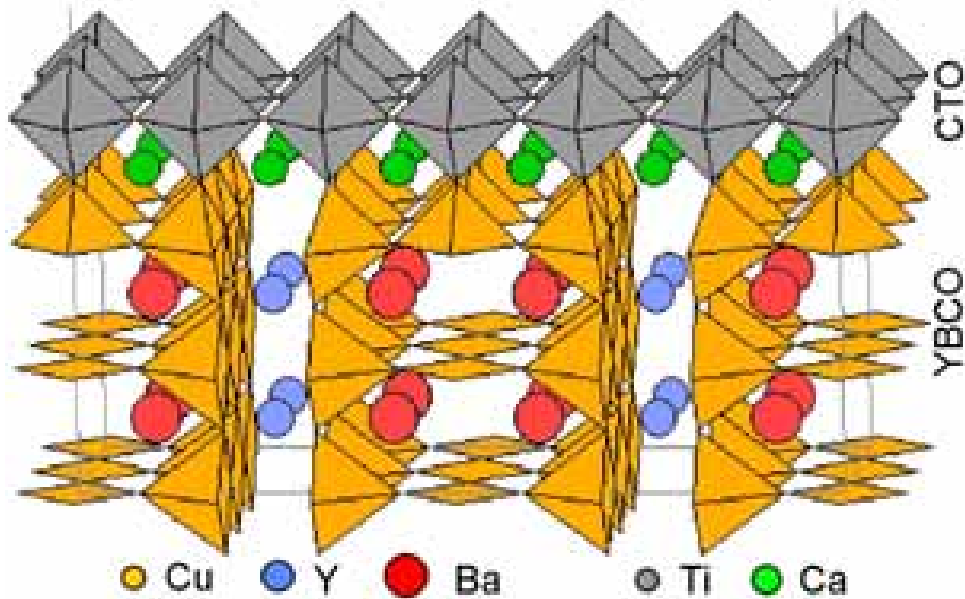

**Fig. 1. Schematic crystal structure at the (100) YBCO/CTO interface, showing oxygen coordination of Cu and Ti atoms. O atoms are present at the vertices of each polyhedron. Note the 5-fold coordination of chain Cu at the interface instead of 4-fold in the bulk, and alignment of $TiO_3$ octahedron along the truncated $CuO_2$ plane edges.**

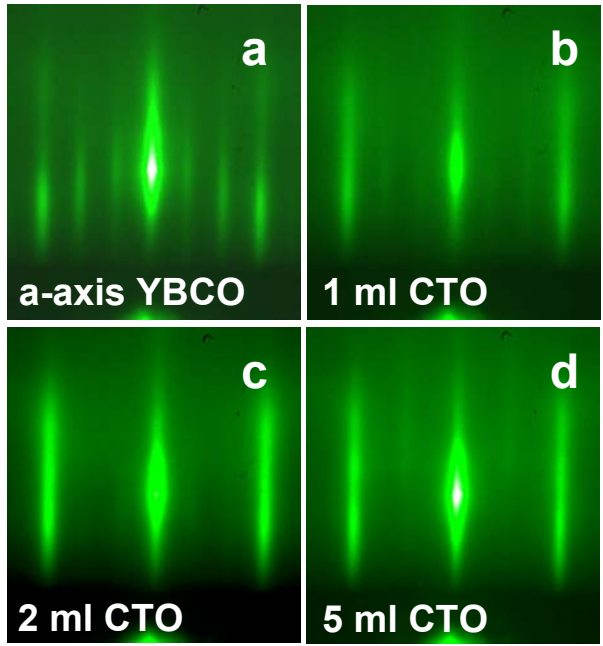

Fig. 2. RHEED images during trilayer deposition. (a) After 2000 Å YBCO; one-third order spots are due to the self-assembled *c*-axis structure lying in the plane of the film. The *a*-axis films are *b-c* twinned, with domain sizes ~100 nm inferred from linewidths of diffraction maxima. (b)-(d) Diffraction during barrier growth.

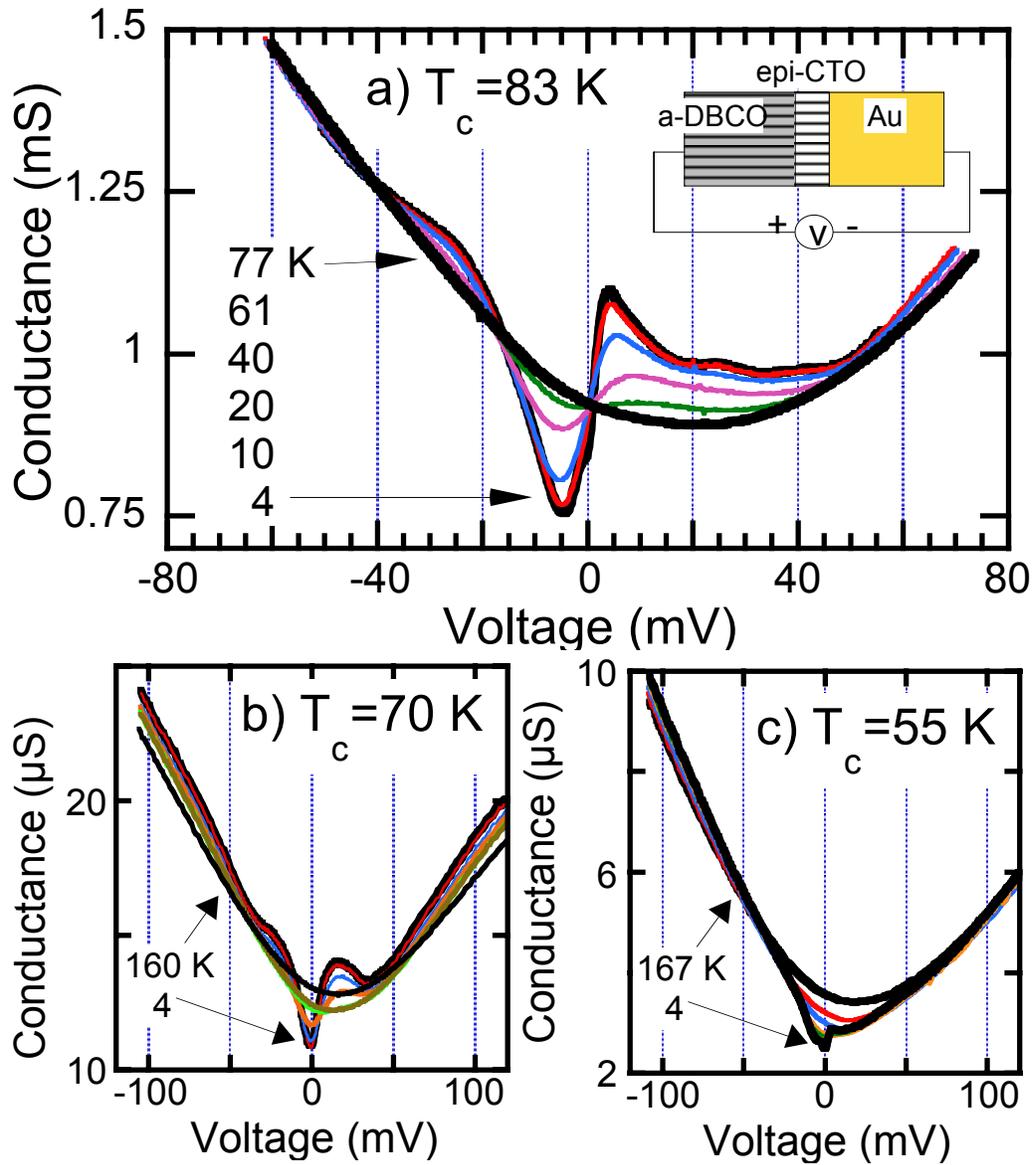

**Fig. 3(a-c). Raw conductance data for 3 NIS junctions with different underdopings, showing temperature evolution of the gap structure for occupied state DoS and filling-in of empty state DoS for $T < T_c$. P-h asymmetry is strongest at highest $T_c$.**

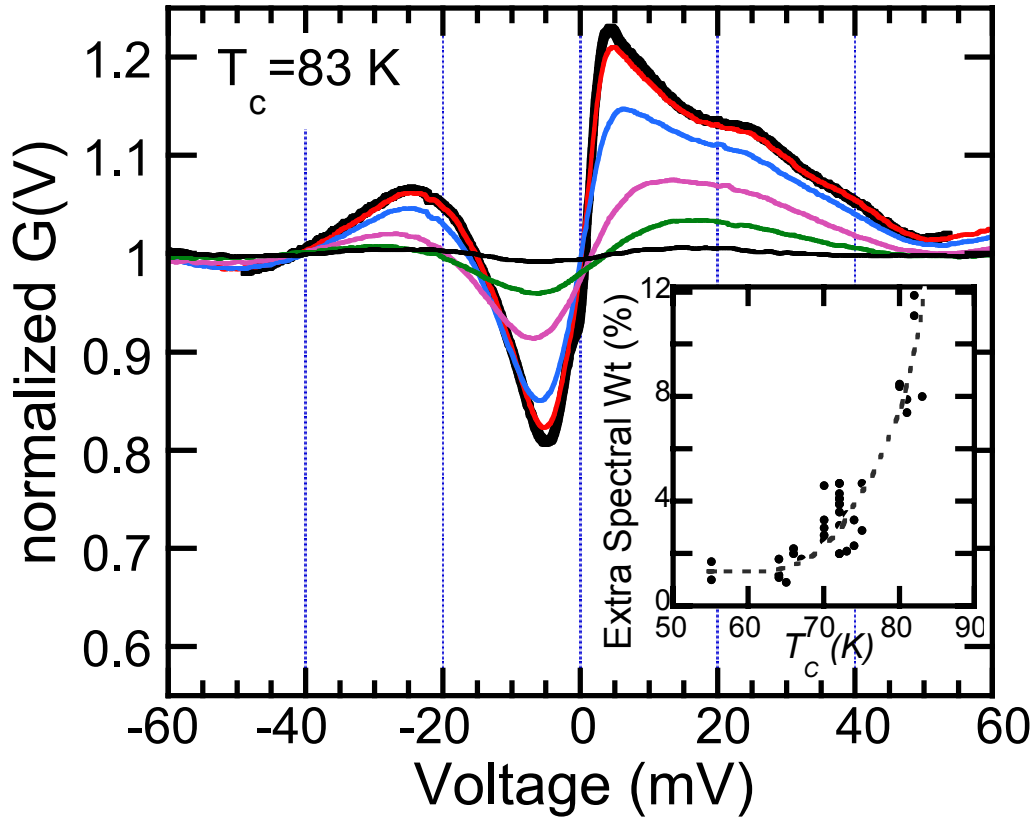

**Fig. 4.** Normalized $G(V)$ data for junction of Fig. 3(a). Inset shows extra spectral weight at $T = 4K$ in empty state DoS versus $T_c$ for all junctions (filled state DoS is conserved). Extra spectral weight decreases with decreasing $T_c$, disappearing as $T_c \to 50K$.